\documentclass[superscriptaddress,aps,prx,reprint,twocolumn,amsmath,amssymb]{revtex4-2}
\usepackage{graphicx}
\usepackage{hyperref}
\usepackage{amssymb}
\usepackage{slashed}
\usepackage{dcolumn}
\usepackage{amsmath}
\usepackage{bm}
\usepackage{colordvi}
\usepackage{algorithm}
\usepackage{algpseudocode}
\usepackage{multirow}
\usepackage{titlesec}
\usepackage{newtxtext,newtxmath}
\usepackage{dsfont}
\usepackage{xcolor}
\usepackage{mathbbol}

\allowdisplaybreaks

\usepackage{mathrsfs}
\makeatletter

\newcommand{\Rmnum}[1]{\expandafter\@slowromancap\romannumeral #1@}
\makeatother

\begin{document}

\title{A tractable framework for phase transitions in phase-fluctuating disordered 2D superconductors: applications to bilayer MoS$_2$ and disordered InO$_x$ thin films}

\author{F. Yang}
\email{fzy5099@psu.edu}

\affiliation{Department of Materials Science and Engineering and Materials Research Institute, The Pennsylvania State University, University Park, PA 16802, USA}

\author{L. Q. Chen}
\email{lqc3@psu.edu}

\affiliation{Department of Materials Science and Engineering and Materials Research Institute, The Pennsylvania State University, University Park, PA 16802, USA}

\date{\today}

\begin{abstract}
  Starting from the purely microscopic model, we go beyond conventional mean-field theory and develop a self-consistent microscopic thermodynamic framework for disordered 2D superconductors. It incorporates the fermionic Bogoliubov quasiparticles, bosonic Nambu–Goldstone (NG) quantum and thermal phase fluctuations in the presence of long-range Coulomb interactions, and topological Berezinskii–Kosterlitz–Thouless (BKT) vortex–antivortex fluctuations on an equal footing, to self-consistently treat the superconducting gap and superfluid density. This unified phase-fluctuating description naturally recovers the previously known limiting results: the superconducting gap in the 2D limit can remain robust against long-wavelength NG phase fluctuations at $T=0^+$ due to Coulomb-induced regularization, while the gradual proliferation of BKT fluctuations as the system approaches criticality drives a separation between the global superconducting transition temperature $T_c$ and the gap-closing temperature $T^*$. In contrast to mean-field theory, which predicts 2D superconductivity to be independent of carrier density and non-magnetic disorder (Anderson theorem), the incorporation of phase fluctuations generates a density- and disorder-dependent zero-point gap $\Delta(0)$ and consequently $T_c$ and $T^*$. Remarkably, applications to bilayer MoS$_2$ [Nat. Nanotechnol.  {\bf 14}, 1123 (2019)] and disordered InO$_x$ thin films [Nat. Phys. {\bf 21}, 104 (2025)] quantitatively reproduce key experimental observations in excellent agreement. The framework offers a useful theoretical tool for understanding phase-fluctuation-dominated superconductivity.

\end{abstract}

\pacs{74.20.-z,74.40.+k,74.62.-c}

\maketitle  

\section{Introduction}

Spontaneous breaking of a continuous symmetry in an ordered quantum phase necessarily gives rise to a gapless collective bosonic excitation, known as the Nambu–Goldstone (NG) mode~\cite{nambu1960quasi,goldstone1961field,goldstone1962broken,nambu2009nobel}. The prediction and identification of such excitations have stimulated extensive research in several subfields of condensed matter physics, including superconductors, charge-density-wave systems, and ferro- and antiferromagnets~\cite{nambu2009nobel}, and have been instrumental in shaping the theoretical foundations that led to the discovery of the Higgs particle in high-energy physics~\cite{anderson1963plasmons,higgs1964broken,englert1964broken}. Understanding how NG excitations govern the stability of ordered phases and shape their critical behaviors has long been a central topic, since this excitation intrinsically acts to restore the broken symmetry, leading to a thermodynamic instability of the phase. For example, this instability forbids the formation of long-range order for two-dimensional (2D) systems, known as the Hohenberg–Mermin–Wagner-Coleman (HMWC) theorem~\cite{hohenberg1967existence,mermin1966absence,coleman1973there}. However, since the isolation of graphene and the subsequent rapid advances in material synthesis and growth, numerous  2D quantum materials are realized, and one of the surprising findings is the observation of robust superconductivity~\cite{saito2016highly,qiu2021recent}, spanning conventional metallic materials in the form of ultrathin films and atomic sheets~\cite{li2023proximity,zhang2010superconductivity,falson2020type,liao2018superconductivity,zhang2015detection,briggs2020atomically}, single-layer transition-metal dichalcogenides such as MoS$_2$~\cite{ye2012superconducting,lu2015evidence,saito2016superconductivity,Costanzo2016}, WTe$_2$~\cite{fatemi2018electrically,sajadi2018gate,song2024unconventional}, and NbSe$_2$~\cite{xu2015experimental,sun2016majorana,wang2017high,tsen2016nature}, monolayer Fe-based~\cite{liu2012electronic,he2013phase,lee2014interfacial,ge2015superconductivity,ding2016high,shi2017enhanced} and Cu-based~\cite{jiang2014high,sterpetti2017comprehensive,yu2019high} high-$T_c$ materials, and magic-angle twisted graphene~\cite{jaoui2022quantum,park2021tunable,cao2018unconventional}.

Despite rapid experimental progress, a unified, quantitative theory describing superconductivity in 2D materials has not been well developed. Specifically, in superconductors, where the $U(1)$ gauge symmetry is spontaneously broken~\cite{nambu1960quasi,nambu2009nobel,schrieffer1964theory}, the associated NG mode corresponds to phase fluctuations of the superconducting order parameter~\cite{ambegaokar1961electromagnetic,nambu1960quasi,nambu2009nobel,littlewood1981gauge}. This excitation in the 3D bulk case is entirely inactive since its original low-energy spectrum is lifted to the high-frequency plasma energy $\omega_p$ by long-range Coulomb interaction via Anderson-Higgs mechanism~\cite{anderson1963plasmons,ambegaokar1961electromagnetic,littlewood1981gauge,yang2019gauge}, rendering phase fluctuations dynamically frozen in the infrared. Consequently, a mean-field description based solely on fermionic Bogoliubov quasiparticles is well-justified in 3D, without the need to explicitly include phase-fluctuation dynamics. In contrast, in the 2D limit the $\omega_p$ becomes gapless, and hence, the phase fluctuations become inevitable. However, the sublinear dispersion $\omega_p\propto\sqrt{q}$ has been shown to avoid the infrared divergence of phase correlations~\cite{PhysRevB.97.054510}, providing a natural route to evade the conventional assumptions underlying the HMWC theorem~\cite{hohenberg1967existence,mermin1966absence,coleman1973there}, which in its original form does not account for the  long-range Coulomb interactions (2D Anderson-Higgs mechanism). But these gapless phase modes remain dynamically active and can generate strong renormalization effects on the zero-point superconducting gap~\cite{PhysRevB.97.054510,yang2021theory,PhysRevB.70.214531,PhysRevB.102.060501}, unlike in 3D. 

The HMWC theorem focuses on the gapless collective NG excitation  that mediates long-wavelength correlations in systems with continuous symmetries, as this excitation provides the smooth-type, long-wavelength collective fluctuation associated with continuous symmetry breaking. However, the 2D superconducting system also supports topological excitations (vortex–antivortex pairs) described by Berezinskii–Kosterlitz–Thouless (BKT) physics~\cite{doi:10.1142/9789814417648_0004,benfatto10,PhysRevB.110.144518}, which arise from the compact nature of the $U(1)$ phase rather than from continuous long-wavelength deformations. These topological excitations, which cannot be continuously connected to the uniform superconducting ground state, drive a universal discontinuous jump in the superfluid stiffness~\cite{PhysRevLett.131.186002,PhysRevB.100.064506}, forcing the superconducting transition temperature $T_c$ below the pairing (gap‐closing) scale $T^*$ and generating a phase‐incoherent pairing regime for $T_c<T<T^*$~\cite{chand2012phase,mondal2011phase,PhysRevB.102.060501}. In this scenario, the superconducting transition at $T_c$ is controlled by loss of global phase coherence rather than by the closing of the pairing gap.  The conventional BKT renormalization-group (RG) procedure takes the bare superfluid stiffness as an external input, typically estimated from Ginzburg–Landau theory~\cite{Halperin1979,benfatto10} or from the clean BCS limit~\cite{PhysRevLett.131.186002,PhysRevB.102.060501}.  This direct input may face challenges~\cite{PhysRevB.107.224502,PhysRevB.99.104509,li2021superconductor,dubi2007nature}, as it does not include renormalization caused by dynamical NG phase fluctuations, and inherits the non-magnetic-disorder-insensitivity of mean-field BCS theory implied by the Anderson theorem~\cite{anderson1959theory,suhl1959impurity,skalski1964properties,andersen2020generalized}, thereby missing the interplay between disorder, phase stiffness, and collective fluctuation effects.   In reality, the 2D materials due to their low dimension are intrinsically disordered, and  experiments performed on superconductors in ultra-thin film and 2D limit demonstrated that increasing disorder concentration can suppress the superconducting transition temperature $T_c$~\cite{mondal2011phase,chockalingam2008superconducting,noat2013unconventional,chand2012phase,fatemi2018electrically,sajadi2018gate,song2024unconventional,sacepe2008disorder}. Numerical Monte Carlo studies~\cite{PhysRevB.96.060508} demonstrate only that that spatial inhomogeneity can smooth out the universal BKT jump in the superfluid stiffness,  a generic feature commonly observed across disordered phase-transition systems. Separately, experiments have  realized that with sufficiently large amount of disorder that drives the superconducting state into a highly granular and spatially inhomogeneous regime, a superconducting state can even transition to an insulating state~\cite{goldman2003superconductor,kowal1994disorder,sacepe2011localization,sherman2015higgs,noat2013unconventional,chand2012phase,sacepe2020quantum}.

Consequently, the 2D superconducting system simultaneously hosts multiple coexisting degrees of freedom, including fermionic Bogoliubov quasiparticles, bosonic NG phase modes, BKT vortex–antivortex topological excitations, as well as the effects of disorder. Existing theoretical approaches typically treat these ingredients separately in the limiting form, while most established numerical frameworks are designed for fermionic quasiparticles and are not capable of handling collective bosonic modes, topological defects, and disorder on an equal and fully self-consistent microscopic footing. Here, we develop a fully microscopic framework that treats these degrees of freedom on equal footing and in a self-consistent manner.  {We incorporate the NG phase fluctuations into the microscopic gap equation in a controlled and gauge-invariant manner and derive a set of self-consistent equations in which the superconducting order parameter and the phase sector mutually renormalize each other while explicitly retaining the
feedback between phase fluctuations and the superconducting gap.  
The phase fluctuations (including BKT physics), the gap equation, and impurity/disorder effects are treated on the same footing. This leads to a single internally consistent set of equations, and hence, a self-consistent beyond-mean-field framework, as well as a computationally tractable scheme for experimentally relevant observables across the corresponding regimes.}

Using this unified theoretical description, the previously known limiting behaviors emerge naturally: the long-range Coulomb interactions convert the gapless NG mode from a linear dispersion $\omega_{\rm NG}(q)\propto q$ to the plasmonic form $\omega_{\rm NG}(q)\propto\sqrt{q}$~\cite{fisher1990presence,fisher1990quantum,mishonov1990plasmon,yang2021theory},
which eliminates infrared divergence and renders the NG phase fluctuation finite at $T=0^+$~\cite{PhysRevB.97.054510,yang2021theory,PhysRevB.70.214531,PhysRevB.102.060501}; independently, the proliferation of BKT fluctuations near criticality drives the decoupling of the global superconducting transition temperature $T_c$ from the pairing scale where the gap closes. Going beyond these limits, we find that Coulomb interactions suppress NG thermal fluctuations, rendering their contribution negligible, while disorder and reduced carrier density strongly amplify NG quantum fluctuations, leading to substantial reduction of the zero-point gap $\Delta(0)$ and consequently the pairing critical temperature $T^*$. The same factors also enhance BKT fluctuations, inducing an explicit density- and disorder-dependent separation $T^*-T_c$.
When applied to gate-tunable bilayer MoS$_2$ superconductors~\cite{Zheliuk2019,Fu2017,Costanzo2016} and to the precursor regime of strongly disordered InO$_x$ films (amorphous superconductors)~\cite{Charpentier2025,sacepe2011localization}, the framework quantitatively reproduces key experimental trends in excellent agreement, including the density- and disorder-dependent suppression of $\Delta(0)$, $T_c$, and $T^*$, showing its practical effectiveness and quantitative reliability for phase-fluctuation-dominated 2D superconductors.

\section{Effective Theory for Phase-Transitions}
\label{secET}

To facilitate readability, we first provide a concise overview of the theoretical framework, which can be directly applied to realistic systems for practical calculations. The full microscopic derivation starting from the microscopic superconducting Hamiltonian within the fundamental path-integral approach and quantum statistic framework is deferred to the following Sec.~\ref{secMD} for completeness. In Sec.~\ref{secPA}, we apply the framework to two representative systems: gate-tunable bilayer MoS$_2$~\cite{Zheliuk2019,Fu2017,Costanzo2016} and strongly disordered amorphous InO$_x$ films~\cite{Charpentier2025,sacepe2011localization}, to demonstrate its applicability to both tunable 2D superconducting platforms and the superconducting precursor regime in disordered systems, respectively.

Specifically, we consider an $s$-wave superconducting system whose order parameter is parameterized as 
\begin{equation}
  \Delta=|\Delta|e^{i\delta\theta({\bf R})}.
\end{equation}
Here, $|\Delta|$ denotes the superconducting gap and $\delta\theta({\bf R})$ is the superconducting phase. The phase itself is not gauge invariant. The physically meaningful, gauge-invariant quantity associated with phase fluctuations is the superfluid momentum, defined as ${\bf p}_s={\bf \nabla_R}\delta\theta({\bf R})/2$, and as established in Ref.~\cite{benfatto10}, its dynamics can be decomposed into two orthogonal components: the longitudinal sector ${\bf p}_{s,\parallel}$ is associated with smooth, long-wavelength phase fluctuations corresponding to the NG mode, whereas the divergence-free transverse sector  ${\bf p}_{s,\perp}$ encodes vortex excitations associated with BKT topology.

The self-consistent gap equation takes the form
\begin{equation}\label{OGE}
 {\frac{1}{U}}=F({p}^2_{s,\parallel},|\Delta|,T)={\sum_{\bf k}}\frac{f(E_{\bf k}^+)-f(E_{\bf k}^-)}{2{E_{\bf k}}},
\end{equation}
where ${{U}}$ is the pairing potential and $f(x)$ is the Fermi distribution function. Here, the quasiparticle spectrum acquires a Doppler shift  ${\bf v_{\bf k}}\cdot{\bf p}_{s,\parallel}$~\cite{fulde1964superconductivity,larkin1965zh,yang2018fulde,yang2018gauge,yang2021theory} and takes the form $E_{\bf k}^{\pm}={\bf v_{\bf k}}\cdot{{\bf p_{s,\parallel}}}\pm{E_{\bf k}}$ where $E_{\bf k}=\sqrt{\xi_{\bf k}^2+|\Delta|^2}$ is the Bogoliubov dispersion,  $\xi_{\bf k}={\hbar^2k^2}/(2m)-\mu$ with $\mu$ being the chemical potential and $m$ the effective mass, and ${\bf v}_{\bf k}={\partial_{\bf k}}\xi_{\bf k}$ is the band velocity. The NG phase fluctuations enter the superconducting gap equation here, playing a role analogous to that of an electromagnetic vector potential (gauge manner)~\cite{ambegaokar1961electromagnetic,nambu1960quasi,nambu2009nobel}.

Since the Doppler shift enters the gap equation symmetrically, Eq.~(\ref{OGE}) depends only on the second moment of the phase fluctuations. Consequently, NG fluctuations contribute through the mean-squared superfluid momentum,
\begin{equation}\label{OPEF}
\langle{p_{s,\parallel}^2}\rangle=S_{\rm th}(T)+S_{\rm zo},
\end{equation}
with the thermal-excitation part
\begin{equation}\label{th}
S_{\rm th}(T)=\int\frac{d{\bf q}}{(2\pi)^2}\frac{2q^2n_B(\omega_{\rm NG})}{D_{q}\omega_{\rm NG}(q)}, 
\end{equation}
and the zero-point oscillation part 
\begin{equation}\label{zo}
S_{\rm zo}=\int\frac{d{\bf q}}{(2\pi)^2}\frac{q^2}{D_{q}\omega_{\rm NG}(q)},  
\end{equation}
where $n_B(x)$ denotes the Bose-Einstein distribution. The NG spectrum follows {\small{$\omega_{\rm NG}(q)=\sqrt{n_sq^2/(D_qm)}$}}, as well established in the  literature~\cite{sun2020collective,yang2021theory,ambegaokar1961electromagnetic,yang2019gauge,PhysRevB.64.140506,PhysRevB.69.184510,PhysRevB.70.214531,PhysRevB.97.054510}. Here, $D_q=2D/(1+2DV_q)$ with  $D$  being the density of states of carriers and {\small{$V_q=2{\pi}e^2/(q\epsilon)$}} the 2D Coulomb potential. The superfluid stiffness is evaluated from the static current-current correlation, yielding
\begin{equation}\label{OSD}
\frac{n_s}{n}=\frac{1}{1+\xi/l}\int{d\xi_k}\int\frac{{d}\theta_{\bf k}}{2\pi}\frac{|\Delta|^2}{2E^3_{\bf k}}[f(E_{\bf k}^-)-f(E_{\bf k}^+)].  
\end{equation}
Here, a prefactor $(1+\xi/l)^{-1}$, where $\xi=\hbar v_F/|\Delta|$ is the superconducting coherence length and $l=v_F\tau$ is the mean free path ($\tau$ is the effective scattering time), is included to incorporate the disorder-induced reduction of phase stiffness while leaving the $s$-wave gap equation intact, consistent with the Anderson-theorem-protected gap~\cite{anderson1959theory,suhl1959impurity,skalski1964properties,andersen2020generalized}. This interpolation traces back to Tinkham's treatment of the London penetration depth $\lambda$ of $s$-wave superconductors~\cite{tinkham2004introduction}, yielding $\lambda^2=\lambda_{\rm clean}^2(1+\xi/l)$ and hence $n_s\rightarrow n_s/(1+\xi/l)$. It reproduces the Mattis–Bardeen dirty-limit result~\cite{yang2024diamagnetic,mattis1958theory} and provides a smooth interpolation between clean and diffusive regimes for finite scattering~\cite{nam1967theory}. This form is further supported by microscopic derivations from multiple established theoretical frameworks, including the Gor'kov formalism~\cite{abrikosov2012methods}, Eilenberger theory~\cite{eilenberger1968transformation,PhysRevB.99.224511,PhysRevLett.25.507}, gauge-invariant kinetic approaches~\cite{yang2018gauge}, and diagrammatic current–current response with vertex corrections~\cite{PhysRevB.99.224511,PhysRevB.106.144509}.

Consequently, Eqs.~(\ref{OGE})–(\ref{OSD}) are solved self-consistently to obtain the temperature-dependent $\Delta(T)$, and in particular, a bare superfluid density $n_s$, which contain fermionic quasiparticle, NG phase-fluctuation, and disorder effects. The bare quantity $n_s$ serves as the initial condition for the standard BKT RG   flow~\cite{benfatto10,PhysRevB.110.144518,PhysRevB.80.214506,PhysRevB.77.100506,PhysRevB.87.184505}:
\begin{equation}\label{OBKT}
\frac{dK}{dL}=-K^2g^2~~~\text{and}~~~\frac{dg}{dL}=(2-K)g,
\end{equation}
with $K(L=0)=\frac{\pi\hbar^2{n_s}}{4mk_BT}$ and $g(L=0)=2\pi{e}^{-c_0K(L=0)}$ (the dimensionless constant $c_0=2/\pi$ in 2D limit). The fully renormalized superfluid density after vortex screening is
\begin{equation}
  {\bar n}_s=\frac{4mk_BT}{\pi\hbar^2}K(L=\infty).
\end{equation}
For phase-transition criticality, the superconducting transition temperature $T_c$ is identified from the BKT criterion (universal jump of the renormalized superfluid stiffness), while the gap-closing temperature is denoted $T^*$, i.e., $\Delta(T^*)=0$.   

As established in numerous experiments~\cite{uemura1989universal,emery1995importance,uemura1991basic,yuli2008enhancement,Charpentier2025,sacepe2011localization}, a key quantity that quantifies the phase rigidity against both NG ($\omega_{\rm NG}\propto\sqrt{n_s/m}$) and BKT ($K(L=0)\propto{n_s/m}$) fluctuations is the superfluid stiffness, defined as
\begin{equation}\label{sstn}
\Theta=\frac{\hbar^2{\bar n}_s}{4m}.  
\end{equation}
A reduction in $\Theta$ softens the NG mode and decreases the initial BKT coupling, thereby amplifying these fluctuations.

 \section{Practical Applications}
\label{secPA}

In the previous section, we summarized the effective mi-
croscopic theory to formulate phase transitions in phase-
fluctuating disordered 2D superconductors, presented in a form
that can be directly applied to realistic systems for practical
calculations. In this section, we directly apply this theoretical framework to two experimentally relevant and conceptually complementary platforms:
(i) gate-tunable bilayer MoS$_2$ \cite{Zheliuk2019,Fu2017,Costanzo2016}, where the carrier density $n$ can be continuously controlled via electrostatic gating, enabling direct access to the density dependence of superconducting properties; and
(ii) strongly disordered amorphous InO$_x$ films \cite{Charpentier2025,sacepe2011localization}, a prototypical system for probing disorder-driven effects in superconductivity. The combination of these two platforms enables independent tuning of carrier density and disorder strength, making them well suited for testing how the zero-temperature pairing gap $\Delta(0)$, the superfluid (phase) stiffness $\Theta$, and the characteristic temperature scales $T^*$ and $T_c$ evolve beyond conventional mean-field expectations.

\subsection{Application to bilayer MoS$_2$}

\begin{figure}[htb]
  \includegraphics[width=8.7cm]{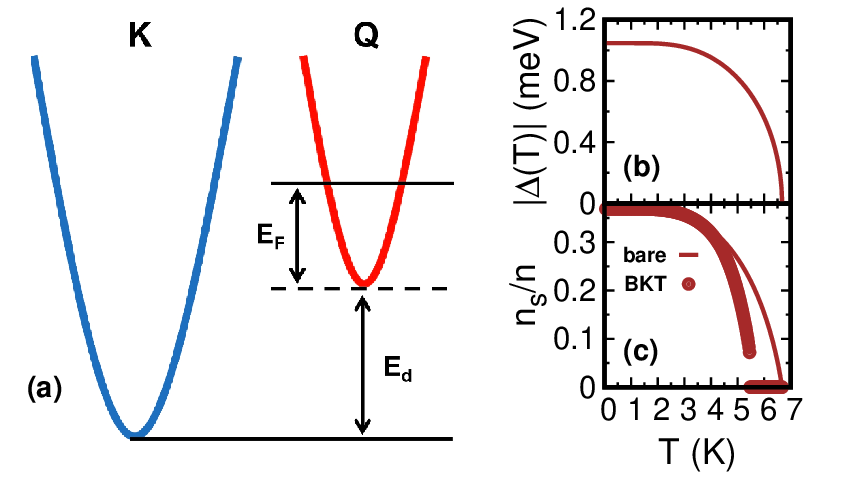}
  \caption{(a) Schematic illustration of the ``lower'' band in $K$ and $Q$ valleys in bilayer MoS$_2$. (b) Temperature dependence of the superconducting gap $|\Delta(T)|$ and (c) the superfluid density, calculated at a carrier density of $n_{\rm 2D}=4.6\times10^{14}~\mathrm{cm^{-2}}$ and a scattering time $\tau\Delta_{\rm MF}=0.85$, chosen to match the results (compared with experiment in Ref.~\cite{Zheliuk2019}) shown in Fig.~\ref{figyc3} and its inset.}
  \label{figyc1}
\end{figure}

Bilayer MoS$_2$ hosts a multi-valley conduction band structure with band minima located at both the $K$ and $Q$ points of the Brillouin zone~\cite{Kormanyos_2015,Xu2014,Wu2013}, as illustrated in Fig.~\ref{figyc1}(a). Experimental estimates indicate that the in-plane electron effective mass is approximately $m_{K}\approx0.5m_{0}$~\cite{PhysRevB.90.045422,Cui2015} for the $K$ valley and $m_{Q}\approx0.6m_0$ for the $Q$ valley~\cite{Cui2015}. Both first-principles calculations and ARPES measurements show that the $Q$-valley band edge lies higher than that of the $K$ valley by an energy offset $E_{d}$ on the order of $\sim80$–$250{\rm meV}$~\cite{Kormanyos_2015}, although the precise value is sensitive to layer stacking, strain, and dielectric environment. The 2H stacking of bilayer MoS$_2$ restores global inversion symmetry, resulting in opposite layer-resolved spin polarization that compensate in momentum space~\cite{Kormanyos_2015,Xu2014,Wu2013,yang2015hole,yang2016spin}. As a result, each valley hosts two effectively spin-degenerate conduction sub-bands (denoted ``lower'' and ``upper''), originating from opposite spin polarization in the two layers.
The energy separation between these two conduction sub-bands in bilayer MoS$_2$, originating from the conduction-band spin splitting in each layer, is negligible at $K$  ($\sim3~\mathrm{meV}$)~\cite{Kormanyos_2015}, but becomes substantially larger at $Q$ ($\sim75~\mathrm{meV}$)~\cite{Kormanyos_2015}. Consequently, defining the lowest $Q$-valley sub-band minimum as the zero of energy, as illustrated in Fig.~\ref{figyc1}(a), the total 2D carrier density is given by
\begin{equation}\label{n2dMOS2}
  n_{\rm 2D}(\mu)\!=\!12D_Q[\mu\!+\!(\mu\!-\!\delta)\theta(\mu\!-\!\delta)]\!+\!8D_K(\mu\!+\!E_d),  
\end{equation}
where $D_{Q(K)}=m_{Q(K)}/(2\pi\hbar^2)$ is the 2D density of states per single valley and single spin at $Q$ ($K$), $\delta$ denotes the energy separation between the upper and lower $Q$-valley sub-bands, and $E_d \equiv E_K - E_Q$ is the energy offset between the lowest $K$-valley and $Q$-valley conduction band minima, and $\theta(x)$ is the step function. The numerical prefactors in Eq.~(\ref{n2dMOS2}) arise from valley and spin multiplicities: $12 = 6(Q\text{ valleys})\times 2(\text{spins})$ and $8 = 2(K\text{ valleys}) \times 2(\text{sub-bands})\times 2(\text{spins})$. 

Theoretical and experimental evidence indicate that superconductivity in gated MoS$_2$ originates predominantly from electrons in the $Q$ valley~\cite{Ding2022,Piatti2018}, likely driven by significantly stronger electron–phonon coupling at $Q/Q'$ compared to $K/K'$ valleys (which exhibit no significant pairing instability). Because electrostatic gating initially populates the $K$-valley states, superconductivity emerges only when the chemical potential $\mu$ is raised above the $Q$-valley band edge. This leads to a well-defined critical carrier density $n_c$, consistent with experimental observations \cite{Ding2022,Piatti2018,Costanzo2016,Fu2017,Zheliuk2019}: for $n < n_c$, carriers occupy only the $K$ valley and the system remains normal, while for $n > n_c$, $Q$-valley occupation enables Cooper pairing and the onset of superconductivity.

We incorporate the effective mass of the $Q$ valley into our framework, and choose the pairing strength $D_Q U = 0.6927$ in our simulations. This yields a superconducting transition temperature of $T_c^{\mathrm{MF}} = 8.9~\text{K}$ in the absence of phase fluctuations (or equivalently, a mean-field zero-point gap $\Delta_{\rm MF}=1.35~$meV), which corresponds to the maximum $T_c$ potentially attainable in the high-density limit inferred from experiments \cite{Zheliuk2019}. We subsequently compute the chemical-potential dependence of the superconducting properties, and convert them to the carrier-density axis using Eq.~(\ref{n2dMOS2}). The band offset is fixed at $E_d = 220~\text{meV}$, in accordance with first-principles estimates for the $Q$-valley band edge \cite{Kormanyos_2015}. The remaining ingredient is the elastic scattering time $\tau$ of the $Q$ valley, which is not directly accessible from experiments and is thus introduced as the sole fitting parameter in our analysis.

\subsubsection{Temperature dependence}

We first examine the temperature dependence. As shown in Fig.~\ref{figyc1}(b), our simulations, which fully incorporate phase fluctuations, find that the superconducting gap remains finite at elevated temperature and displays a continuous, gradual-suppression thermal evolution, confirming that it is not destroyed by thermal NG phase fluctuations. This robustness originates from the extremely small thermal population of NG modes, controlled by the phase-fluctuation phase-space factor $q^2 n_B(\omega_{\rm NG})$ in Eq.~(\ref{th}). Crucially, in the presence of long-range Coulomb interactions, the NG dispersion transforms from the acoustic-like form $\omega_{\rm NG}(q)\propto{q}$ to a plasmonic form $\omega_{\rm NG}(q)\propto\sqrt{q}$. This crossover dramatically reshapes the low-energy phase space, restricting thermally excitable NG modes to a narrow momentum range while endowing them with a much higher propagation velocity compared to the original linear mode. As a consequence, the thermal occupation of long-wavelength phase fluctuations is strongly suppressed, making their contribution to thermodynamic quantities and critical phase fluctuations negligible. As established in Refs.~\cite{PhysRevB.97.054510,yang2021theory,yang2025efficient}, this mechanism provides a natural means to circumvent the infrared divergence of phase correlations at $T\neq 0$ that encodes in the HMWC theorem \cite{hohenberg1967existence,mermin1966absence,coleman1973there}.

The superfluid density exhibits a temperature evolution characteristic of BKT physics~\cite{doi:10.1142/9789814417648_0004,benfatto10,PhysRevB.110.144518,PhysRevB.100.064506,PhysRevLett.131.186002,PhysRevB.102.060501},  as shown in Fig.~\ref{figyc1}(c). With increasing temperature, gradually emerging BKT fluctuations lead to a strong renormalization of the superfluid density, causing the renormalized ${\bar n}_s$ to decay much faster than the bare one $n_s$. This eventually leads to a universal, discontinuous jump of the superfluid density and drives the superconducting transition at $T_c$. The $T_c$ lies far below the temperature $T^{*}$, opening an intermediate phase-disordered pseudogap regime ($T_c < T < T^*$) characterized by phase-incoherent pairing, where Cooper pairs remain formed but lack global phase coherence, and therefore do not exhibit zero resistivity.

\subsubsection{Carrier-density dependence}

\begin{figure}[htb]
  {\includegraphics[width=8.8cm]{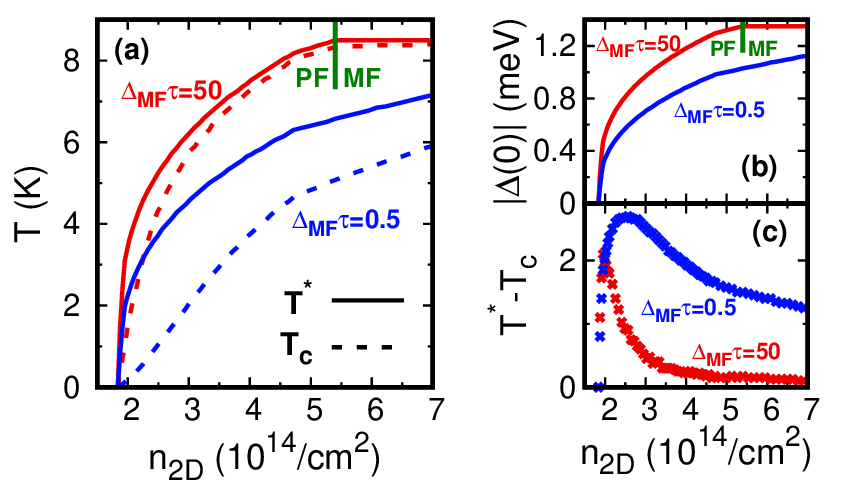}}
  \caption{Carrier-density dependence of (a) the pair-formation temperature $T^{*}$ and the superconducting transition temperature $T_c$; (b) the zero-point gap $|\Delta(0)|$; and (c) the width of the finite-temperature pseudogap regime, defined as $T^{*}-T_c$, shown for both the clean regime ($\Delta_{\rm MF}\tau=50$) and the dirty regime ($\Delta_{\rm MF}\tau=0.5$). The solid green lines in panels (a) and (b) mark the boundary between the mean-field (MF) regime and the phase-fluctuation(PF)–dominated regime. }
\label{figyc2}
\end{figure}

We next focus on the carrier-density dependence, plotted in Fig.~\ref{figyc2}. As seen from the figures, in the high-density limit ($n_{\rm 2D}>5.4\times10^{14}\mathrm{cm^{-2}}$) and in the weak-scattering regime ($\Delta_{\rm MF}\tau=50\gg1$), all fluctuation effects are minimal. In this regime, the superconducting transition temperature $T_c$ closely tracks the pair-formation scale $T^{*}$ [red crosses in Fig.~\ref{figyc2}(c)], indicating only marginal BKT renormalization. Both the zero-point gap $|\Delta(0)|$ [red curve in Fig.~\ref{figyc2}(b)] and the characteristic energy scales $T^{*}$ and $T_c$ [red curves in Fig.~\ref{figyc2}(a)] show negligible dependence on carrier density, signaling a fluctuation-free regime where pairing and phase coherence collapse onto a single energy scale governed by mean-field theories. Within this high-density region, as indicated by the blue curves/crosses in Fig.~\ref{figyc2}(a)-(c), only strong disorder can drive the system into a fluctuation-dominated regime, an effect discussed later.

Upon decreasing the carrier density from the high-density regime, regardless of the disorder strength, both the superconducting transition temperature $T_c$ and the pair-formation temperature $T^{*}$ decrease systematically during carrier depletion. The reduction of $T^{*}$ closely tracks the suppression of the zero-point gap $|\Delta(0)|$ [shown in Fig.~\ref{figyc2}(b)], which exactly originates from the renormalization induced by NG zero-point oscillations (quantum fluctuations) [Eq.~(\ref{zo})]. In contrast, BKT phase fluctuations primarily control the separation between the two energy scales, causing the pseudogap window, defined as $T^{*}-T_c$, to broaden as the carrier density decreases, as shown in Fig.~\ref{figyc2}(c).  All of these trends stem from the reduction of the superfluid stiffness $\Theta$ in Eq.~(\ref{sstn}), which softens the NG collective mode and weakens the bare BKT coupling, thereby enhancing phase-fluctuation effects and driving stronger renormalization with decreasing density, beyond mean-field expectations. 
 
All superconducting characteristics, including the pairing temperature $T^{}$, the transition temperature $T_c$, the zero-point gap $|\Delta(0)|$, and the pseudogap window $T^{*}-T_c$, collapse precipitously and vanish as the system approaches the critical carrier density $n_{\rm 2D}\rightarrow{n_c}+0^+$. This behavior arises from the aforementioned valley-selective nature of the pairing~\cite{Ding2022,Piatti2018}: superconductivity is predominantly supported by the $Q$ valleys, while the $K$ valleys remain non-superconducting. As the carrier density decreases, the population in the $Q$ valleys diminishes, directly driving the collapse of the superconducting energy scales, as the system no longer hosts the electrons that participate in superconducting pairing.

\begin{figure}[htb]
  {\includegraphics[width=6.6cm]{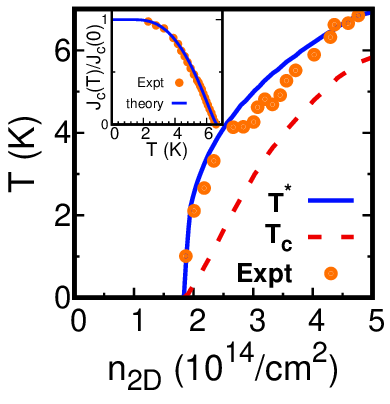}}
 \caption{Comparison between theory and experiment. The scattering time $\tau$ is the only fitting parameter, determined by matching the pair-formation temperature $T^{*}$ at $n_{\rm 2D}=4.76\times10^{14}\mathrm{cm^{-2}}$ to the experimental nominal transition temperature $T^*_{\rm e}=6.87~$K in Ref.~\cite{Zheliuk2019}. The resulting density dependence of $T^{*}$ and $T_c$ shows good agreement with experiment without additional fitting parameters. Inset: the calculated temperature dependence of the critical current compared with the experimental data~\cite{Zheliuk2019}. In the calculation of the critical current, a vector potential ${\bf A}$ is introduced through a Doppler shift ${\bf v}_{\bf k}\cdot{\bf A}$ in the quasiparticle spectrum. The critical vector potential ${\bf A}_c(T)$ is determined self-consistently at each temperature, and the critical current is obtained as $J_c(T)\propto n_s(T) A_c(T)$, showing quantitative agreement with the experimental critical current extracted from differential resistance measurements \cite{Zheliuk2019}.}    
\label{figyc3}
\end{figure}

Having established the underlying physical trends, we now compare directly with experiment, treating the elastic scattering time $\tau$ as the only adjustable parameter. In the measurements reported in Ref.~\cite{Zheliuk2019}, the superconducting transition temperature was defined operationally as the temperature at which the resistance drops to 50\% of its normal-state value. This criterion places the experimentally extracted transition point between $T_c$ and $T^{*}$, but closer to $T^{*}$, consistent with the picture that pairing forms prior to the establishment of long-range phase coherence. By fitting the scattering time $\tau$ to match the experimental nominal transition temperature $T^*_{\rm e}$ at a representative carrier density of $n_{\rm 2D}=4.76\times10^{14}~\mathrm{cm^{-2}}$, we obtain the theoretical density dependence of both $T^{*}$ and $T_c$, shown in Fig.~\ref{figyc3}. As seen from the figure, without introducing any additional fitting parameters, the calculated $T^{*}$ exhibits good qualitative agreement with the experimental trend across the density range. In particular, at $n_{\rm 2D}=4.6\times10^{14}~\mathrm{cm^{-2}}$, the experimentally extracted pairing onset temperature is $T^{*}_{\rm e}\approx6.7~\mathrm{K}$, while the zero-resistance temperature inferred from the measured resistance curve in Ref.~\cite{Zheliuk2019} is approximately $6~\mathrm{K}$. These values are in close agreement with our theoretical predictions of $T^{*}=6.7~\mathrm{K}$ and $T_c=5.75~\mathrm{K}$. Furthermore, at this same carrier density, without any additional parameter tuning, the calculated temperature dependence of the critical current quantitatively reproduces the experimental behavior extracted from the differential resistance data \cite{Zheliuk2019}, as shown in the inset of Fig.~\ref{figyc3}, underscoring the predictive capability of the framework.

\subsection{Application to disordered InO$_x$ thin films: disorder dependence}

Disordered InO$_x$ thin films constitute one of the most extensively studied platforms for disorder-tuned quantum phase transitions in superconductors~\cite{PhysRevB.89.035149,sacepe2020quantum,sacepe2011localization,Dubouchet2019,PhysRevB.109.144501,Charpentier2025,PhysRevB.91.220508}, where structural and electronic disorder can be continuously tuned via oxygen stoichiometry or post-annealing. Tunneling spectroscopy reveals a characteristic two-stage evolution: upon cooling below a pairing-onset temperature $T^*$, a pseudogap gradually opens in the density of states (inset of Fig.~\ref{figyc4}(a), data from Ref.~\cite{sacepe2011localization}), signaling the formation of preformed Cooper pairs without global phase coherence. Upon further cooling to $T_c$, global phase stiffness emerges through a BKT-type jump~\cite{Charpentier2025}, establishing long-range superconducting coherence. In this regime, as shown in Fig.~\ref{figyc4}(a) (data from Ref.~\cite{Charpentier2025}), increasing disorder strongly suppresses the superfluid stiffness while a finite local pairing gap persists, indicating that the transition at $T_c$ is driven primarily by phase fluctuations rather than by gap closing. These features make disordered InO$_x$ an ideal platform for testing theories of disorder-driven effects in phase-fluctuating superconductivity.

\begin{widetext}
\begin{center}  
  \begin{figure}[htb]
  {\includegraphics[width=17.7cm]{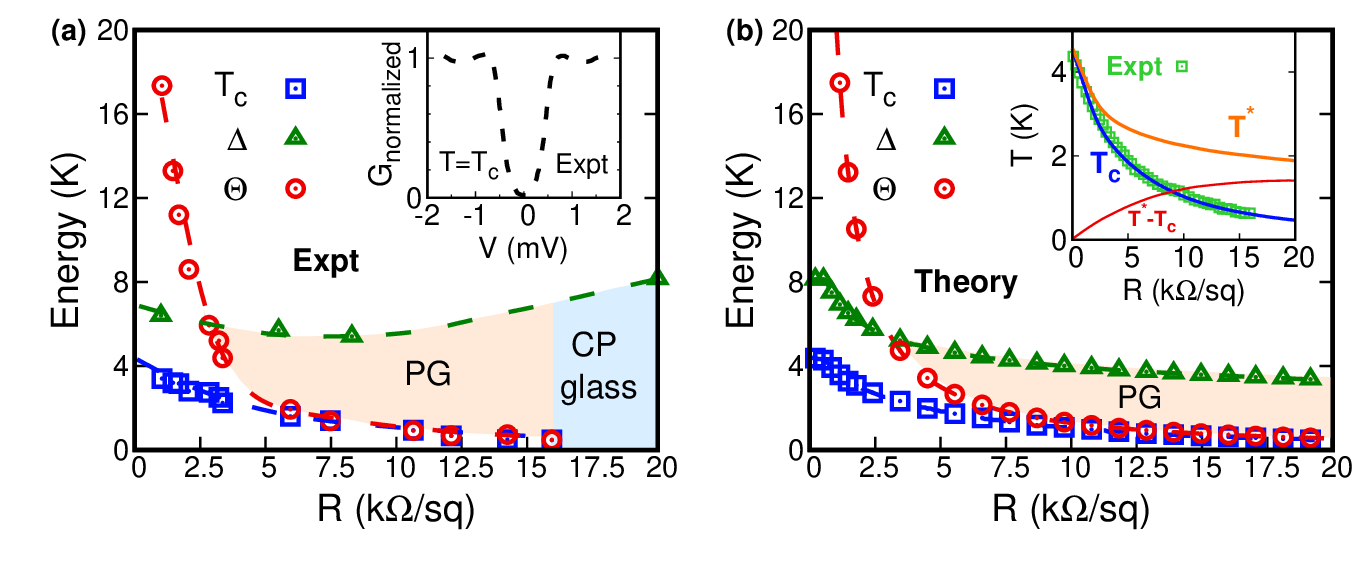}}
  \caption{(a) Experimentally measured and (b) theoretically calculated disorder dependence of the zero-point gap $|\Delta(0^+)|$ and the superfluid stiffness $\Theta(0^+)$ (in units of K) as well as the critical temperature $T^*$ of superconducting InO$_x$ thin film. Experimental data come from Ref.~\cite{Charpentier2025}. The inset of (a) displays the experimentally measured tunneling spectroscopy {(obtained by local scanning tunneling microscopy (STM) measurements of the normalized differential conductance, which reflects the opening of the superconducting gap)} for a sample with $R\sim5~\mathrm{k\Omega/sq}$ at $T=T_c=1.7~\mathrm{K}$, showing a pseudogap in the density of states that persists up to $\sim 6~\mathrm{K}$~\cite{sacepe2011localization}. In our simulations, we approximate the $0^+$ limit with $T = 25~\mathrm{mK}$, corresponding to the lowest experimentally accessible temperature~\cite{Charpentier2025}. The inset of (b) shows the disorder dependence of the theoretically calculated temperatures $T^*$, $T_c$, and $T^* - T_c$ (solid curves), while the squares represent the corresponding experimental values from panel (a), showing that the theoretical and experimental $T_c$ curves nearly coincide.} 
\label{figyc4}
  \end{figure}
\end{center}
\end{widetext}

In our simulations, we set $DU = 0.409$, which yields a mean-field zero-point gap $\Delta_{\rm MF} = 0.7~\mathrm{meV}$ in the absence of phase fluctuations~\cite{Charpentier2025}. The effective mass is taken as $m = 0.35,m_0$, consistent with experimental and first-principles estimates for amorphous InO$_x$~\cite{BUCHHOLZ2013475,Buchholz2014InOx}, and the chemical potential is set to $\mu = 0.2~\mathrm{eV}$. Although experimental InO$_x$ samples are thin films rather than strictly 2D, the superconducting gap equation at fixed $DU$ depends only weakly on dimensionality. In contrast, phase fluctuations are strongly enhanced in the quasi-2D geometry relevant to these films. We therefore treat phase fluctuations within a 2D framework, in line with the experimentally proposed picture of a phase-driven superconducting transition~\cite{Charpentier2025}, while taking the dimensionless prefactor $c_0$ entering the BKT initial condition [Eq.~(\ref{OBKT})] as the sole fitting parameter, determined by matching the experimentally measured $T_c$ at sheet resistance $R = 7.5~\mathrm{k\Omega/sq}$~\cite{Charpentier2025}.

The sheet resistance in our simulation is approximately expressed as $R\propto m/(ne^2\tau)$. Then, the self-consistently calculated disorder dependence of the gap $|\Delta(0^+)|$ and the superfluid stiffness $\Theta(0^+)$ at $T=25~\mathrm{mK}$, together with the critical temperatures $T^*$ and $T_c$, is presented in Fig.~\ref{figyc4}(b). Remarkably, by direct comparison between Fig.~\ref{figyc4}(a) and Fig.~\ref{figyc4}(b), we find that for $R<15~\mathrm{k\Omega/sq}$ (prior to entering the insulating regime, discussed later), the theoretical results quantitatively reproduce the experimental trends with excellent agreement.

Specifically, in the clean limit $R\rightarrow{0}$, the superfluid stiffness $\Theta(0^+)$ (circles in Fig.~\ref{figyc4}) far exceeds the pairing scale $|\Delta(0^+)|$ (triangles in Fig.~\ref{figyc4}), indicating a phase-stiff superconducting state with $T_c \sim T^*$ [inset of Fig.~\ref{figyc4}(b)], where the transition is primarily controlled by Cooper-pair formation. As disorder increases within the weak-disorder regime $R<5~\mathrm{k\Omega/sq}$, the superfluid stiffness $\Theta(0^+)$ is rapidly suppressed by impurity scattering via the renormalization factor $1/(1+\xi/l)$ in Eq.~(\ref{OSD}). This strengthens the zero-point NG phase fluctuations [Eq.~(\ref{zo})], which in turn suppress both the zero-point gap $|\Delta(0^+)|$ (Fig.~\ref{figyc4}) and the pairing-onset temperature $T^*$ [inset of Fig.~\ref{figyc4}(b)] beyond the mean-field expectation. Furthermore, the reduction of $\Theta(0^+)$ also amplifies the BKT phase fluctuations, causing an increasing separation between $T^*$ and $T_c$ [inset of Fig.~\ref{figyc4}(b)], accompanied by a continued decrease in $T_c$ with increasing disorder, consistent with experimental observations~\cite{PhysRevB.89.035149,sacepe2020quantum,sacepe2011localization,Dubouchet2019,PhysRevB.109.144501,Charpentier2025,PhysRevB.91.220508}. The resulting hierarchy reversal, $\Theta(0^+)<|\Delta(0^+)|$, signals the onset of a significant pseudogap regime ($T_c<T<T^*$) in which preformed Cooper pairs exist without global phase coherence.

Upon further increasing disorder into the strong-scattering regime $R \in [5, 15]~\mathrm{k\Omega/sq}$, both theoretical calculations [Fig.~\ref{figyc4}(b)] and experimental measurements [Fig.~\ref{figyc4}(a)] find $\Theta(0^+) \approx k_B T_c$ (circles and triangles) over a wide disorder range, indicating that the superconducting transition at $T_c$ becomes fully governed by phase fluctuations. Notably, in this regime, the effects of both NG and BKT fluctuations remain active but their net contribution to gap suppression and $T^*-T_c$ narrows, and as a result, the superconducting gap shown in Fig.~\ref{figyc4}(b) and the separation $T^* - T_c$ shown in its inset exhibit weak disorder dependence, in contrast to the pronounced disorder sensitivity observed in the weak-scattering regime.

With further increasing disorder to $R > 15~\mathrm{k\Omega/sq}$, experiments~\cite{Charpentier2025} show that the system evolves into an insulating state, characterized by a sharp upturn in low-temperature resistance, while both $\Theta$ and $T_c$ collapse abruptly, signaling the onset of a disorder-driven (likely first-order) superconductor–insulator transition. Remarkably, however, the local tunneling gap does not vanish but persists and even increases inside the insulating phase, as shown in Fig.~\ref{figyc4}(a).  This regime has been interpreted experimentally as a Cooper-pair glass insulator~\cite{Charpentier2025}, where localized Cooper pairs form a glassy state described by a spin-glass–type order parameter~\cite{10.21468/SciPostPhys.17.2.066}. At present, such a strongly localized regime lies beyond the scope of our microscopic theory, which is formulated within a diffusive transport framework and does not incorporate disorder-induced localization or spatial inhomogeneities.  {Specifically, the present work focuses on phase fluctuations at long length scales, while incorporating disorder solely through its scattering effect, placing the system in the diffusive regime. This description is well justified for weak to moderate disorder, where the superconducting state remains spatially homogeneous and the contribution from amplitude fluctuations is secondary. This approximation is based on the fact that at low temperatures the excitation energy associated with phase fluctuations, corresponding to the gapless NG mode, is significantly lower than that of amplitude fluctuations, which are associated with a gapped Higgs mode. As a result, phase fluctuations constitute the dominant low-energy degrees of freedom and are naturally the primary focus of the present study.  In the regime of strong disorder, however, spatial inhomogeneity and fluctuations of the pairing amplitude become essential. In such a regime, amplitude fluctuations alone can lead to a separation and inversion of relevant energy scales~\cite{PhysRevLett.81.3940}, and the inclusion of phase fluctuations may further drive the system toward a non-superconducting (insulating or glassy) state.  A systematic extension of the present microscopic theory of phase-fluctuating superconductivity to explicitly incorporate localization physics~\cite{PhysRevB.32.5658} on equal footing and self-consistently renormalize the superfluid density from first principles, while achieving quantitative agreement with experiments in the localization-dominated regime, remains an important open theoretical challenge. Our approach is therefore limited to the diffusive regime and does not extend into the localized regime where strong inhomogeneity dominates.}

Nevertheless, experiments indicate that the emergence of the Cooper-pair glass insulator occurs via a {\sl first-order} transition, reflecting a competition between two distinct minima in the free-energy landscape: one associated with a diffusive, phase-fluctuating superconducting state, and the other corresponding to a localized, insulating glassy configuration of Cooper pairs.  This first-order separation effectively protects the stability of each phase, suppressing the mutual interplay that would otherwise renormalize their respective properties. Thus, in light of the experimentally established first-order boundary between the diffusive superconducting and Cooper-pair–glass regimes, we expect that the conclusions drawn and the results obtained within the diffusive regime remain robust and quantitatively reliable before the system enters the fully localized limit.

\section{Microscopic Derivation}
\label{secMD}

In Sec.~\ref{secET}, we summarized the effective microscopic theory to formulate phase transitions in phase-fluctuating disordered 2D superconductors, presented in a form that can be directly applied to realistic systems for practical calculations. 
  For theoretical completeness, in this section, we provide its microscopic origin and detailed derivations starting from the underlying superconducting Hamiltonian.

\subsection{Model Hamiltonian}

We begin with the standard microscopic model of the $s$-wave superconductors, described by the Hamiltonian~\cite{schrieffer1964theory}:
\begin{eqnarray}
  H_0&=&\int\!\!{d{\bf x}}\Big[\sum_s\psi^{\dagger}_s({\bf x})\xi_{\bf {\hat p}}\psi_s({\bf x})-U\psi^{\dagger}_{\uparrow}({\bf x})\psi^{\dagger}_{\downarrow}({\bf x})\psi_{\downarrow}({\bf x})\psi_{\uparrow}({\bf x})\Big]\nonumber\\
  &&\mbox{}+\frac{1}{2}\int{d{\bf x}d{\bf x}'}V({\bf x}-{\bf x}')\rho^{\dagger}({\bf x})\rho({\bf x}').~~\label{Ham}
\end{eqnarray}
Here, $\psi_s({\bf x})$ is the fermionic field operator with spin $s=\uparrow,\downarrow$,  $\xi_{\bf {\hat p}}={\bf {\hat p}^2}/(2m)-\mu$ with ${\hat {\bf p}}=-i\hbar{\bf \nabla}$ being the momentum operator and $\mu$ the chemical potential, and $U$ is the local pairing potential. The last term describes the long-range Coulomb interaction $V({\bf x-x'})$, with density operator $\rho({\bf x})=\sum_{s}\psi^{\dagger}_{s}({\bf x})\psi_{s}({\bf x})$

The action of this model can be determined based on the Hamiltonian (\ref{Ham}), and written as
\begin{align}
  &S[\psi,\psi^{\dagger}]=\int\!\!\!{dx}\Big[\sum_{s}\psi^{\dagger}_s(x)(i\partial_{x_0}\!-\!\xi_{\hat {\bf p}})\psi_s(x)+\!U\psi^{\dagger}_{\uparrow}(x)\psi^{\dagger}_{\downarrow}(x)\nonumber\\
    &~\mbox{}\times\psi_{\downarrow}(x)\psi_{\uparrow}(x)\Big]\!-\!\frac{1}{2}\int{dxdx'}V(x-x')\rho^{\dagger}(x)\rho(x'),
\end{align}
with the four-vector ${x}=(x_0,{\bf x})$. Then, applying the standard Hubbard-Stratonovich transformation introduces the superconducting order parameter $\Delta(x)=|\Delta(x)|e^{i\delta\theta(x)}$ as well as the  Coulomb-interaction-related Hartree field $\mu_H(x)$~\cite{schrieffer1964theory,yang2021theory,yang2024diamagnetic,yang2023optical,sun2020collective}, yielding
\begin{align}
  &S[\psi,\psi^{\dagger}]=\int\!\!\Big[\sum_{s}{dx}\psi^{\dagger}_s(x)[i\partial_{x_0}\!-\!\xi_{\hat {\bf p}}\!-\!\mu_H(x)]\psi_s(x)\!-\!\frac{|\Delta(x)|^2}{U}\nonumber\\
  &~~\mbox{}-\psi^{\dagger}(x){\hat \Delta}(x)\psi(x)\Big]\!+\!\frac{1}{2}\int\!\!\!{dx_0}d{\bf q}\frac{|\mu_H(x_0,{\bf q})|^2}{V_{\bf q}}.
\end{align}
Here, $\psi(x)=[\psi_{\uparrow}(x),\psi^{\dagger}_{\downarrow}(x)]^T$ represents the field operator in Nambu space, and ${\hat \Delta}(x)=\Delta(x)\tau_++\Delta^*(x)\tau_-$ with $\tau_i$ being the Pauli matrices in Nambu space~\cite{abrikosov2012methods}. Further applying the unitary transformation $\psi(x){\rightarrow}e^{i\tau_3\delta\theta(x)/2}\psi(x)$, one obtains
\begin{align}
  &S=\!\!\int\!\!{dx}\Big\{\sum_{s}\psi^{\dagger}_s(x)\Big[i\partial_{x_0}-\frac{\partial_{x_0}\delta\theta(x)}{2}-\xi_{{\hat {\bf p}}+{\bf p}_s}-\mu_H(x)\Big]\psi_s(x)\nonumber\\
    &~\mbox{}-\!\psi^{\dagger}(x)|{\Delta}(x)|\tau_1\psi(x)\!-\!\frac{|\Delta(x)|^2}{U}\Big\}\!+\!\int\!\!\!{dx_0}d{\bf q}\frac{|\mu_H(x_0,{\bf q})|^2}{2V_{\bf q}},\label{AA1}
\end{align}
where ${\bf p}_s={\bf \nabla}\delta\theta/2$. We then introduce the center-of-mass coordinate $[R=\frac{x_1+x_2}{2}=(t,{\bf R})]$ and relative coordinate $[r=x_1-x_2=(r_0,{\bf r})]$ for the pairing electrons, and perform a Wigner transformation that maps ${\bf R}\rightarrow{\bf q}$ and ${\bf r}\rightarrow{\bf k}$. Within the imaginary-time path integral approach, through the standard integration over the fermi field and the Hartree field, the effective action of the gap and phase fluctuation can be written as~\cite{schrieffer1964theory,yang2021theory,yang2024diamagnetic,yang2023optical,sun2020collective}  (See Sec.~\ref{EA} for details)
\begin{align}
  {S}_{\rm eff}=&\int{d{R}}\Big\{{\sum_{p_n,{\bf k}}}\ln[(ip_n-E_{\bf k}^+)(ip_n-E_{\bf k}^-)]\!-\!\frac{|\Delta|^2}{U}-\frac{np_s^2}{2m}\Big\}\nonumber\\
  &\mbox{}+\frac{1}{2}\int{dt}\sum_{\bf q}\frac{2D}{1\!+\!2DV_q}\Big(\frac{\partial_{t}\delta\theta_{\bf q}}{2}\!\Big)^2,\label{action}
\end{align}
with Doppler-shifted quasiparticle dispersion $E_{\bf k}^{\pm}={\bf v_{\bf k}}\cdot{{\bf p_s}}\pm{E_{\bf k}}$ and $p_n=(2n+1)\pi{T}$ fermionic Matsubara frequency. Here, the Bogoliubov
dispersion $E_{\bf k}=\sqrt{\xi_{\bf k}^2+|\Delta|^2}$ and ${\bf v}_{\bf k}={\partial_{\bf k}}\xi_{\bf k}$.

\subsection{Derivation of the effective theory}

The effective action in Eq.~(\ref{action}), which encodes both temporal (compressibility) and spatial (superfluid stiffness) responses, describes not only the equilibrium state of the superconducting condensate but also its collective modes and long-wavelength hydrodynamic regime. Within this framework, the coupled dynamics of the superconducting amplitude (gap) and phase fluctuations can be systematically derived.

\subsubsection{Gap equation}

Stationarity with respect to the gap yields the self-consistent gap equation:
\begin{eqnarray}
  0&=&\frac{|\Delta|}{U}+\sum_{p_n,{\bf k}}\frac{|\Delta|}{(ip_n-E^+_{\bf k})(ip_n-E^-_{\bf k})}\nonumber\\
  &=&\frac{|\Delta|}{U}+\sum_{{\bf k}}\frac{|\Delta|}{2E_{\bf k}}[f(E_{\bf k}^+)-f(E_{\bf k}^-)],
\end{eqnarray}
which can be written in the compact form
\begin{equation}
 {\frac{1}{U}}=F({p}^2_{s},|\Delta|,T)={\sum_{\bf k}}\frac{f(E_{\bf k}^+)-f(E_{\bf k}^-)}{2{E_{\bf k}}}.
\end{equation}
The structure of this equation is analogous to that appearing in finite–center-of-mass–momentum pairing theories such as the Fulde–Ferrell–Larkin–Ovchinnikov (FFLO) framework~\cite{fulde1964superconductivity,larkin1965zh,yang2018fulde,yang2018gauge,yang2021theory}. Here, inversion and angular symmetries enforce that the Doppler field enters only through $p_s^2$ after momentum averaging, ensuring that the pairing sector depends solely on the magnitude of the superfluid momentum.

\subsubsection{NG and BKT separation}

As established in Ref.~\cite{benfatto10}, the phase-fluctuation field naturally decomposes (Helmholtz decomposition) into longitudinal and transverse sectors:
 \begin{equation}
 {\bf p}_s={\bf p}_{s,\parallel}+{\bf p}_{s,\perp},~~\text{with}~~~{\bm \nabla}\times{\bf p}_{s,\parallel}=0~~~\text{and}~~~{\bm \nabla\cdot}{\bf p}_{s,\perp}=0.
\end{equation} 
Only the longitudinal component ${\bf p}_{s,\parallel}$ (curl-free) enters the global gap equation and controls uniform pair-breaking via NG fluctuations. In contrast, the transverse component ${\bf p}_{s,\perp}$ (divergence-free) affects superconductivity indirectly by renormalizing the superfluid stiffness through BKT vortex physics, without modifying the spatially averaged gap.

{\sl Longitudinal sector (NG mode).---}The longitudinal sector ${\bf p}_{s,\parallel}$ 
corresponds to the gapless smooth-type, long-wavelength NG phase fluctuations~\cite{ambegaokar1961electromagnetic,nambu1960quasi,nambu2009nobel,littlewood1981gauge}. Specifically, through the Euler-Lagrange equation of motion with respect to the long-wavelength phase fluctuation, i.e., $\partial_{\mu}[\frac{\partial{S_{\rm eff}}}{\partial(\partial_{\mu}\delta\theta/2)}]=\partial_{\delta\theta/2}S_{\rm eff}$, in frequency and momentum space, the equation of motion of the NG phase fluctuations reads
\begin{equation}
\Big(\frac{n_sq^2}{m}-D_q\omega^2\Big)\frac{\delta\theta(\omega,{\bf q})}{2}=0,  \label{PEF}  
\end{equation}
which can be written in the compact form
\begin{equation}\label{EOMPS}
[\partial_t^2+\omega^2_{\rm NG}(q)]{\bf p}_{s,\parallel}({\bf q})=0,  
\end{equation}
where the NG-mode spectrum $\omega^2_{\rm NG}(q)=n_sq^2/(D_qm)$. 

The associated superfluid stiffness is determined by 
\begin{align}
  \frac{n_s{\bf p_s}}{m}\!=&\frac{n{\bf p}_s}{m}\!+\!{\sum_{p_n{\bf k}}}\frac{2{\bf k}(ip_n\!-\!{\bf k}\!\cdot\!{\bf p}_s/m)}{m(ip_n\!-\!E_{\bf k}^+)(ip_n\!-\!E_{\bf k}^-)}\nonumber\\
  =&\frac{{\bf p}_sk_F^2}{m^2}{\sum_{\bf k}}\Big[\frac{\partial_{E_{\bf k}}A_{\bf k}}{2}-\partial_{\xi_{\bf k}}\Big(\frac{\xi_{\bf k}A_{\bf k}}{2E_{\bf k}}\Big)\Big]\nonumber\\
  =&\frac{{\bf p}_sk_F^2}{m^2}\sum_{\bf k}\frac{|\Delta|^2}{E_{\bf k}}\partial_{E_{\bf k}}\Big(\frac{A_{\bf k}}{2E_{\bf k}}\Big)\!\approx\!-\frac{{\bf p}_sk_F^2}{m^2}|\Delta|^2\sum_{\bf k}\frac{A_{\bf k}}{2E^3_{\bf k}},~~~~ 
\end{align}
with $A_{\bf k}=f(E_{\bf k}^+)-f(E_{\bf k}^-)$. The superfluid density then takes the form
\begin{eqnarray}
  n_s=2\mu|\Delta|^2\sum_{\bf k}\frac{f(E_{\bf k}^-)-f(E_{\bf k}^+)}{2E^3_{\bf k}}.  \label{SD}
\end{eqnarray}
Equation (\ref{EOMPS}) reproduces the standard NG mode dispersion derived in gauge-invariant superconducting electrodynamics and microscopic treatments~\cite{ambegaokar1961electromagnetic,nambu1960quasi,littlewood1981gauge,yang2021theory,sun2020collective}. The corresponding superfluid density also coincides with results obtained from Gor'kov theory, kinetic formulations, and current-current correlation approaches~\cite{abrikosov2012methods,yang2018gauge,yang2019gauge,yang2024diamagnetic,sun2020collective}. Since the phase fluctuations originate from the collective NG bosons associated with spontaneous $U(1)$ symmetry breaking~\cite{nambu1960quasi,goldstone1961field,goldstone1962broken}, their statistical averaging must be carried out within a quantum-statistical framework. The mean-square phase fluctuation can be obtained either by directly computing the bosonic propagator in the Matsubara representation~\cite{yang2021theory,PhysRevB.102.014511,abrikosov2012methods}, or, equivalently, from a stochastic formulation employing a thermal field ${\bf J}_{\rm th}(\omega,{\bf q})$ that obeys the fluctuation-dissipation theorem~\cite{Landaubook}:
\begin{equation}
\langle{J}_{\rm th}(\omega,{\bf q}){J}^*_{\rm th}(\omega',{\bf q}')\rangle=\frac{(2\pi)^3\gamma\omega\delta(\omega-\omega')\delta({\bf q-q'})}{\tanh(\beta\omega/2)},   
\end{equation}
and $\gamma=0^+$ is a phenomenological damping constant. The dynamics of the phase from Eq.~(\ref{PEF}) is then given by 
\begin{equation}
\Big(\frac{n_sq^2}{2m}-\frac{D_q}{2}\omega^2+i\omega{\gamma}\Big)\frac{\delta\theta({\omega,{\bf q}})}{2}={J}_{\rm th}(\omega,{\bf q}).    
\end{equation}  
Then, defining $D(\omega,{\bf q})=D_q/2[\omega^2-\omega^2_{\rm NG}({\bf q})\big]-i\omega\gamma$, the average of the phase fluctuations is given by
\begin{eqnarray}
  \langle{p_{s,\parallel}^2}\rangle\!&=&\!\int\frac{d\omega{d\omega'}d{\bf q}d{\bf q'}}{(2\pi)^6}\frac{({\bf q}\cdot{\bf q'})\langle{J}_{\rm th}(\omega,{\bf q}){J}^*_{\rm th}(\omega',{\bf q}')\rangle}{D(\omega,{\bf q})D(\omega',{\bf q}')}\nonumber\\
  &=&\!\!\int\frac{d\omega{d{\bf q}}}{(2\pi)^3}\frac{4q^2\gamma\omega\coth(\beta\omega/2)}{\big[D_q\big(\omega^2\!-\!\omega^2_{\rm NG}({\bf q})\big)\big]^2+4\omega^2\gamma^2},\label{EEEE}
\end{eqnarray}
yielding 
\begin{equation}\label{BET}
{p}^2_{s,\parallel}=\langle{p_{s,{\rm NG}}^2}\rangle=\int\frac{d{\bf q}}{(2\pi)^2}\frac{q^2[2n_B(\omega_{\rm NG})+1]}{D_{q}\omega_{\rm NG}(q)},  
\end{equation}
which accounts for thermal Bose excitations (described by Bose distribution function $n_B$) and quantum zero-point fluctuations of the NG mode.

{\sl Transverse sector (BKT physics).---}Only the transverse sector ${\bf p}_{s,\perp}$ contains the singular (vorticity-carrying) phase texture and encodes the BKT physics~\cite{benfatto10}. This follows from the quantized circulation of the phase gradient,
\begin{equation}
\oint{\bf p}_s\cdot{d{\bf l}}=\pi\sum_jq_j,~~~q_j\in\mathbb{Z},
\end{equation}
which can be expressed as 
\begin{equation}
\oint{\bf p}_s\cdot{d{\bf l}}=\int_S({\bm \nabla}\times{\bf p_s})\!\cdot\!{d{\bf s}}=\int_S({\bm \nabla}\times{\bf p_{s,\perp}})\!\cdot\!{d{\bf s}}. 
\end{equation}
Equivalently, the transverse sector can be expressed as a set of singular vorticity sources, ${\bm \nabla}\times{\bf p_{s,\perp}}=\pi\sum_jq_j\delta({\bf r-r_j})$, representing vortices and antivortices as point-like topological defects located at ${\bf r_j}$, each carrying an integer vorticity (topological charge) $q_j$.  

These topological defects strongly disrupt global phase coherence and renormalize the superfluid stiffness. The related BKT RG description has been well established in the literature. It takes the bare superfluid density $n_s$ as input to to the universal RG equations~\cite{benfatto10,PhysRevB.110.144518,PhysRevB.80.214506,PhysRevB.77.100506,PhysRevB.87.184505}:
\begin{equation}\label{BKT1}
\frac{dK}{dL}=-K^2g^2~~~\text{and}~~~\frac{dg}{dL}=(2-K)g,
\end{equation}
with initial conditions
\begin{equation}
K(L=0)=\frac{\pi}{k_BT}\frac{\hbar^2{n_s}}{4m}~~~\text{and}~~~g(L=0)=2\pi{e}^{-\mu_v(T)/(k_BT)}.
\end{equation}
Here, $\mu_v$ is the vortex–core energy, which can be estimated from the condensation energy lost within a core of radius $L_v=\hbar{v_F}/(\pi|\Delta|)$~\cite{benfatto10}. In the 2D limit, one has  
\begin{align}
\frac{\mu_v}{k_BT}&=\frac{\pi{L^2_v}E_{c}}{k_BT}\frac{n_s}{n}=\pi\frac{\hbar^2v_F^2}{\pi^2|\Delta|^2}\frac{D|\Delta|^2}{2k_BT}\frac{n_s}{n}=\frac{2}{\pi^2}\frac{\pi\hbar^2n_s}{4m}\nonumber\\
&=c_0K(L=0),  \label{BKT2}
\end{align}
where $E_{c}={D|\Delta|^2}/{2}$ is the condensation-energy density.

Integrating the RG flow to $L\rightarrow\infty$ yields the renormalized superfluid density \begin{equation}
  {\bar n}_s=\frac{4mk_BT}{\pi\hbar^2}K(L=\infty),
\end{equation}
which incorporates the screening from BKT fluctuations. The separatrix $2-\pi{K}=0$ marks the Nelson–Kosterlitz universal jump~\cite{benfatto10,PhysRevB.110.144518,PhysRevB.80.214506}. Trajectories with  $K<2/\pi$ flow to the disordered phase where $K(L=\infty)\rightarrow0$ and $g\rightarrow\infty$, signaling vortex proliferation and loss of phase coherence,  while those with $K>2/\pi$ approach a finite stiffness $K(L=\infty)$ with $g\rightarrow0$, characteristic of a phase in which vortices bind into neutral vortex–antivortex pairs.

\subsubsection{Inclusion of disorder}

For isotropic $s$-wave pairing, non-magnetic disorder impacts the amplitude and phase sectors in qualitatively different ways. The superconducting gap equation remains unchanged in the diffusive limit, consistent with Anderson’s theorem~\cite{anderson1959theory,suhl1959impurity,skalski1964properties,andersen2020generalized} (see Sec.~\ref{ED}). In contrast, the superfluid density is determined by the current–current response kernel~\cite{PhysRevB.99.224511,PhysRevB.106.144509}, where vertex corrections due to impurity scattering necessarily appear, and becomes consequently suppressed by elastic impurity scattering, as realized by various theoretical approaches using Gor'kov theory~\cite{abrikosov2012methods}, Eilenberger transport~\cite{eilenberger1968transformation,PhysRevLett.25.507,PhysRevB.99.224511}, gauge-invariant kinetic equations~\cite{yang2018gauge}, and diagrammatic formulations incorporating vertex corrections to the current–current correlation~\cite{PhysRevB.99.224511,PhysRevB.106.144509}. For practical modeling without solving full impurity-dressed kernels, we adopt the well-established clean-to-dirty interpolation proposed by Tinkham~\cite{tinkham2004introduction}, which relates the penetration depth as $\lambda^2=\lambda_{\rm clean}^2(1+\xi/l)$ and hence $n_s\rightarrow n_{s,{\rm clean}}/(1+\xi/l)$, since the Tinkham's approach recovers the Mattis–Bardeen dirty-limit behavior~\cite{mattis1958theory,yang2024diamagnetic}, connects smoothly to finite-scattering extensions~\cite{nam1967theory}, and quantitatively captures penetration-depth measurements across disorder regimes in conventional superconductors~\cite{PhysRevB.76.094515,PhysRevB.72.064503,PhysRevB.10.1885,PhysRevB.6.2579}. The superfluid density in our framework then takes the form
 \begin{equation}\label{SD2}
\frac{n_s}{n}=\frac{1}{1+\xi/l}\int{d\xi_k}\int\frac{{d}\theta_{\bf k}}{2\pi}\frac{|\Delta|^2}{2E^3_{\bf k}}[f(E_{\bf k}^-)-f(E_{\bf k}^+)],
 \end{equation}
where $\xi=\hbar v_F/|\Delta|$ is the coherence length and $l=v_F\tau$ is the mean free path  with $\tau$ being the effective scattering time. This result is consistent with  the Mattis–Bardeen formula in the dirty limit, where the penetration depth (or equivalently the superfluid stiffness) depends explicitly on the scattering rate through~\cite{mattis1958theory,tinkham2004introduction} 
\begin{equation}
1/\lambda^2(0)\propto|\Delta|\sigma_n,     
\end{equation}
explicitly linking disorder (via the normal-state electrical conductivity $\sigma_n=ne^2\tau/m$ with ) to the loss of superfluid rigidity, 
\begin{equation}
\frac{n_s(T=0)}{n}\propto\frac{l}{\xi}, 
\end{equation}
in the dirty-limit $(\xi\gg{l})$ regime.

From the expressions above, it is clear that the disorder dependence entering our superfluid stiffness remains entirely within the diffusive regime, where quasiparticles retain a finite mean free path and transport is governed by impurity-modified coherence rather than Anderson localization. In particular, the superfluid response varies smoothly with the ratio $l/\xi$ and contains no signatures of the exponentially suppressed diffusion constant, or the strong spatial inhomogeneity characteristic of the localization-dominated regime. Consequently, all disorder effects incorporated in our framework correspond to impurity-dressed diffusive transport, and the theory does not capture the physics of the Cooper-pair–glass phase, which necessarily involves strong localization of paired states and the breakdown of diffusive quasiparticle motion.\\

\section{Summary}

We develop a microscopic theoretical framework for disordered 2D superconductors, designed to capture the phase-fluctuating superconductivity. It treats fermionic quasiparticles, bosonic NG phase modes, long-range Coulomb interactions,  topological BKT vortex fluctuations and the scattering effects on equal footing, and hence, self-consistently incorporates (i) zero-point and thermal NG phase fluctuations in the pairing gap equation, (ii) disorder-induced suppression of the superfluid stiffness $\Theta$, and (iii) vortex-driven renormalization of the physical stiffness $\bar{n}_s(T)$ via BKT RG flow. This enables simultaneous evaluation of the superconducting gap, superfluid density, the pairing scale $T^{*}$, and the global phase-coherence temperature $T_c$. Several key physical consequences emerge naturall. First, long-range Coulomb interactions convert the NG dispersion from $\omega_{\rm NG}\propto q$ to a plasmonic form $\omega\propto\sqrt{q}$~\cite{fisher1990presence,fisher1990quantum,mishonov1990plasmon,yang2021theory}, strongly reducing the thermally accessible phase space and protecting the gap against thermal NG phase fluctuations. Second, zero-point NG fluctuations renormalize the zero-point gap $\Delta(0)$ when the superfluid stiffness is reduced, leading to a strong density- and disorder-dependent suppression of $T^{*}$, a phenomenon beyond conventional mean-field or Anderson-theorem~\cite{anderson1959theory,suhl1959impurity,skalski1964properties,andersen2020generalized} descriptions. Third, BKT vortex fluctuations suppress phase coherence much more efficiently than pairing~\cite{benfatto10,PhysRevB.110.144518,PhysRevB.80.214506,PhysRevB.77.100506,PhysRevB.87.184505}, producing an extended pseudogap regime $T_c < T < T^{*}$ whose width increases as carrier density decreases or disorder increases.

We apply the framework to gated bilayer MoS$_2$~\cite{Kormanyos_2015} and strongly disordered InO$_x$ thin films~\cite{Charpentier2025}. The simulations quantitatively reproduce key experimental trends and explain multiple recent observations in phase-fluctuation-dominated superconductors. Our results provide a unified, parameter-efficient microscopic description bridging pairing-dominated and phase-fluctuations-dominated regimes,  and yields quantitative predictions for $T^{*}$, $T_c$, $\Delta(0)$, and $\Theta(0)$, including their scaling with carrier density and disorder. It further identifies experimentally testable signatures, such as stiffness-controlled gap renormalization, BKT-driven pseudogap expansion, and non-mean-field suppression of superconductivity. Most importantly, the framework is highly practical. It relies on only a small set of physical parameters, connects directly to measurable quantities, and can be readily deployed for quantitative modeling of various 2D superconducting materials.\\

{\it Acknowledgments.---}
  This work is supported by the US Department of Energy, Office of Science, Basic Energy Sciences, under Award Number DE-SC0020145 as part of Computational Materials Sciences Program.  F.Y. and L.Q.C. also appreciate the generous support from the Donald W. Hamer Foundation through a Hamer Professorship at Penn State.

\begin{widetext}
  
\begin{appendix}

\section{Derivation of effective action}
\label{EA}

In this part, we present the detailed derivation of the effective action governing gap and phase dynamics [Eq.~(\ref{action})]. Starting from the fermionic action of Eq.~(\ref{AA1}), introducing the center-of-mass coordinate $[R=\frac{x_1+x_2}{2}=(t,{\bf R})]$ and relative coordinate $[r=x_1-x_2=(r_0,{\bf r})]$ for the pairing electrons~\cite{schrieffer1964theory,yang2021theory,yang2024diamagnetic,yang2023optical,sun2020collective,yang2019gauge,abrikosov2012methods}, focusing on the global (macroscopic) gap of the system (i.e., neglecting the spatial fluctuations of the gap amplitude), one obtains
\begin{align}
  S[\psi,\psi^{\dagger}]=&\int{dx}\psi^{\dagger}(x)\big[i\partial_{\tau}\!-\!{\bf p}_s\cdot{\bf v_{\bf {\hat k}}}\!-\!\xi_{\hat {\bf k}}\tau_3\!-\!|\Delta|\tau_1\!+\!\Sigma(R)\tau_3\big]\psi(x)-\!\int{dR}\big[\frac{|\Delta|^2}{U}\!+\!\Sigma(R)\big]\!+\!\frac{1}{2}\int{dt}d{\bf q}\frac{|\mu_H({\bf q},x_0)|^2}{V_q}\nonumber\\
  =&\int{dx}\psi^{\dagger}(x)\Big\{(i\partial_{\tau}-{\bf p}_s\cdot{\bf v_{\bf {\hat k}}}-\xi_{\hat {\bf k}}\tau_3-|\Delta|\tau_1+\Sigma(R)\tau_3\Big\}\psi(x)-\!\int{dR}\Big(\frac{p_s^2}{2m}+\frac{|\Delta|^2}{U}\Big)+\!\frac{1}{2}\int{dt}d{\bf q}\frac{|\mu_H({\bf q},x_0)|^2}{V_q},
\end{align}
in which the self-energy $\Sigma=\frac{p_s^2}{2m}+\mu_H(R)+\frac{\partial_t\delta\theta(R)}{2}$, and we have used the expansion of the kinetic energy, 
\begin{equation}
\xi_{{\hat {\bf p}}+{\bf p}_s}=\xi_{\bf {\hat k}}+{\bf p}_s\cdot{\bf v_{\hat k}}+{\bf p}_s^2/(2m), 
\end{equation}
as well as the relation
\begin{equation}
\int{dt}{d{\bf R}}\Big[\mu_H(R)+\partial_t\delta\theta(R)/2\Big]=0. 
\end{equation}
Integrating out the fermion fields generates the functional~\cite{schrieffer1964theory,yang2021theory,yang2024diamagnetic,yang2023optical,sun2020collective}
\begin{eqnarray}
S_{\rm eff}=\int{dR}\Big({\rm {\bar T}r}\ln{G_0^{-1}}-\sum_n^{\infty}\frac{1}{n}{\rm {\bar T}r}\{[\Sigma(R)\tau_3G_0]^n\}\Big)-\int{dR}\Big(\frac{p_s^2}{2m}+\frac{|\Delta|^2}{U}\Big)+\frac{1}{2}\int{dt}d{\bf q}\frac{|\mu_H({\bf q},x_0)|^2}{V_q},
\end{eqnarray}
where the fermionic Green function in the Matsubara representation and momentum space is given by 
\begin{equation}\label{GreenfunctionMr}
G_0(p)=\frac{ip_n\tau_0-{\bf p}_s\cdot{\bf v_k}\tau_0+\xi_{\bf k}\tau_3+|\Delta|\tau_1}{(ip_n-E_{\bf k}^+)(ip_n-E_{\bf k}^-)},  
\end{equation}  
with the four-vector $p=(ip_n, {\bf k})$.  Retaining the lowest two orders (i.e., $n=1$ and $n=2$) gives 
\begin{equation}
  S=\int{dR}\Big\{\sum_{p_n,{\bf k}}\ln[(ip_n-E_{\bf k}^+)(ip_n-E_{\bf k}^-)]-{\tilde \chi}_3\frac{p_s^2}{2m}+\chi_{33}\Big[\frac{p_s^2}{2m}+\mu_H({R})+\frac{\partial_t\delta\theta(R)}{2}\Big]^2\!-\!\frac{|\Delta|^2}{U}\Big\}+\frac{1}{2}\int{dt}d{\bf q}\frac{|\mu_H({\bf q},x_0)|^2}{V_q},
\end{equation}  
with the correlation coefficients
\begin{eqnarray}
{\tilde \chi}_3&=&\sum_{\bf k}\Big[1+\sum_{p_n}{\rm Tr}[G_0(p)\tau_3]\Big]=\sum_{\bf k}\Big[1+\frac{\xi_{\bf k}}{E_{\bf k}}\Big(f(E_{\bf k}^+)-f(E_{\bf k}^-)\Big)\Big]=-\frac{k_F^2}{2m}\sum_{\bf k}\partial_{\xi_{\bf k}}\Big[\frac{\xi_{\bf k}}{E_{\bf k}}\Big(f(E_{\bf k}^+)-f(E_{\bf k}^-)\Big)\Big]=n, \\
\chi_{33}&=&-\frac{1}{2}\sum_p{\rm Tr}[G_0(p)\tau_3G_0(p)\tau_3]=-\sum_{p_n,{\bf k}}\frac{(ip_n-{\bf p}_s\cdot{\bf v_k})^2+\xi_{\bf k}^2-|\Delta|^2}{(ip_n-E_{\bf k}^+)^2(ip_n-E_{\bf k}^-)^2}=-\sum_{\bf k}\partial_{\xi_{\bf k}}\Big[\frac{\xi_{\bf k}}{2E_{\bf k}}\Big(f(E_{\bf k}^+)-f(E_{\bf k}^-)\Big)\Big]=D.~~~~~
\end{eqnarray}
Integrating out the Hartree field produces the effective action of the gap and phase fluctuation: 
\begin{equation}
 S_{\rm eff}=\int{dR}\Big\{\sum_{p_n,{\bf k}}\ln[(ip_n-E_{\bf k}^+)(ip_n-E_{\bf k}^-)]-\frac{np_s^2}{2m}-\frac{|\Delta|^2}{U}\Big\}+\int{dt}d{\bf q}\frac{D}{1+2DV_q}\Big(\frac{\partial_{t}\delta\theta_{\bf q}}{2}+\frac{p_s^2}{2m}\Big)^2,
\end{equation}
and neglecting higher-order phase interactions recovers Eq.~(\ref{action}).

\section{Self-consistent treatment of zero-point oscillations and integral cutoff}

The zero-point oscillations of the bosonic NG mode induce a finite expectation value of $S_{\rm zo}$ [Eq.~(\ref{BET})] and formally yields a nonzero $\langle p_{s,\parallel}^2 \rangle$ at $T=0$, which would incorrectly imply $n_s(T=0)<n$ in Eq.~(\ref{SD}) and violate Galilean invariance.  As discussed in Refs.~\cite{PhysRevB.64.140506,PhysRevB.69.184510}, this inconsistency for superfluid density can be resolved by introducing a shift of the chemical potential.

A similar conceptual issue is well recognized beyond superconductivity. In first-principles lattice-dynamical calculations, zero-point lattice vibration does not explicitly appear in thermal phonon populations. There, its effect is absorbed into renormalized model parameters~\cite{verdi2023quantum,wu2022large,yang2024microscopic}, such that finite-temperature vibrations (phonon excitations) are described exclusively by the Bose distribution $n_B(x)$. This separation between vacuum fluctuations and finite-temperature excitations is also standard in quantum field theory, where the vacuum is understood as a dynamically fluctuating state rather than an empty reference point, and where temperature-independent contributions are absorbed into ground-state parameters via normal ordering~\cite{peskin2018introduction}. Such a treatment preserves fundamental constraints including Ward identities and sum-rule conservation, and has proven quantitatively reliable in strongly quantum paraelectric systems~\cite{verdi2023quantum,muller1979srti,li2019terahertz,cheng2023terahertz,wu2022large}. Guided by this principle, we isolate the zero-temperature quantum phase motion from finite-temperature collective fluctuations. Specifically, the zero-point contribution $S_{\rm zo}$ [Eq.~(\ref{SD})] is absorbed into a redefined pairing interaction, i.e., by renormalizing $U$ to $\tilde U$, such that the quantities $\tilde U$ and $\langle{p_{s,\parallel}^2}\rangle=S_{\rm th}$ replace $U$ and $\langle{p_{s,\parallel}^2}\rangle=S_{\rm th}+S_{\rm zo}$ in the framework, respectively. The renormalized coupling is fixed through
\begin{equation}\label{S1}
\frac{1}{\tilde U}={F[p_{s,\parallel}^2=0,|\Delta(T=0)|,T=0]},
\end{equation}
while the physical zero-point gap $|\Delta(T=0)|$ incorporates the full zero-point contribution via
\begin{equation}\label{S2}
\frac{1}{U}={F[p_{s,\parallel}^2=S_{\rm zo},|\Delta(T=0)|,T=0]}.
\end{equation}
This construction ensures that the essential zero-point renormalization on gap remains unchanged, in consistency with the previous studies~\cite{PhysRevB.97.054510,yang2021theory,PhysRevB.70.214531,PhysRevB.102.060501} and, importantly, that the superfluid density satisfies $n_s(T=0)\equiv n$ in the clean limit,  consistent with Galilean invariance and prior analyses~\cite{PhysRevB.64.140506,PhysRevB.69.184510}.

With this vacuum contribution removed from the dynamical sector, finite-temperature phase fluctuations enter solely through the thermal component,
\begin{equation}\label{S3}
\frac{1}{\tilde U} = F\left[p_{s,\parallel}^2 = S_{\rm th}\big(T,n_s(T)\big), |\Delta(T)|, T \right],
\end{equation}
ensuring that thermal and quantum fluctuations play distinct and non-overlapping roles in the gap equation. Within this formulation, all quantities evolve continuously between $T=0$ and $T=0^+$ limit, and the resulting framework remains internally consistent while eliminating unphysical zero-point suppression of superfluid stiffness.

In the framework, the momentum cutoff in the NG-mode bosonic integrals is set to $q_{\rm cutoff}\sim\xi^{-1}$, consistent with Refs.~\cite{yang2021theory,PhysRevB.97.054510,PhysRevB.70.214531}.    This choice reflects the fact that quasiparticle scattering limits phase coherence on length scales shorter than the superconducting coherence length. Consequently, NG phase fluctuations with wavelengths shorter than the superconducting coherence length are strongly suppressed and make a negligible contribution to thermodynamic properties.

For the BCS shell, we set the energy cutoff to $E_{\rm cutoff}=2\Delta_{\rm MF}$ by default. This is a phenomenological, data-aligned choice that ensures consistency between the quasiparticle gap scale and the phase-fluctuation sector.  This cutoff is not rigid and can be refined on a material-by-material basis as more data become available (for example, using quasiparticle gaps from tunneling/ARPES, the width of the specific-heat jump, band-structure anisotropy, disorder levels, or strong-coupling corrections) to tighten or relax $E_{\rm cutoff}$. We have checked that reasonable variations of this cutoff do not alter our main conclusions.

\section{Anderson theorem}
\label{ED}

To facilitate the reading and understanding of how impurity effects enter the gap and stiffness channels asymmetrically, we reproduce here the well-known Anderson theorem~\cite{anderson1959theory,suhl1959impurity,skalski1964properties,andersen2020generalized}. Starting from the standard Gor'kov equation in Matsubara presentation:
\begin{equation}\label{Dyson}
[ip_n-H_0({\bf k})-\Sigma(ip_n,{\bf k})]G(ip_n,{\bf k})=1,
\end{equation}
where the self-energy from nonmagnetic scattering is written as 
\begin{equation}
\Sigma(ip_n,{\bf k})=c_i\sum_{{\bf k}'}V_{\bf kk'}\tau_3G(ip_n,{\bf k}')\tau_3V_{\bf k'k},  
\end{equation}
with $V_{\bf kk'}$ being the electron-impurity interaction, and the Bogoliubov–de Gennes Hamiltonian $H_0({\bf k})=\xi_{\bf k}\tau_3+|\Delta|\tau_1$, one finds that the Green function retains the analytical structure, yielding a
 renormalized equation as~\cite{anderson1959theory,suhl1959impurity,skalski1964properties,andersen2020generalized}
\begin{equation}
ip_n-H_0(k)-\Sigma(ip_n,{\bf k})=i{\tilde p}_n-{\tilde H}_0(k)=i{\tilde p}_n-\xi_{\bf k}\tau_3-|{\tilde \Delta}|\tau_1.\label{SEM_renormalization}  
\end{equation}
This leads to a solution from Eq.~(\ref{Dyson}):
\begin{equation}
G(ip_n,k)=\frac{i{\tilde p}_n\tau_0+\xi_{\bf k}\tau_3+|{\tilde \Delta}|\tau_1}{(i{\tilde p}_n)^2-\xi_{\bf k}^2-|{\tilde \Delta}|^2}. \label{SEM_renormalizedGreenfunction}
\end{equation}
Substituting Eq.~(\ref{SEM_renormalizedGreenfunction}) to Eq.~(\ref{SEM_renormalization}), one finds 
\begin{eqnarray}
  {\tilde p}_n&=&p_n+\Gamma_0\frac{{\tilde p}_n}{\sqrt{({\tilde p}_n)^2+|\Delta|^2}},\\
  |{\tilde \Delta}|&=&|\Delta|+\Gamma_0\frac{|{\tilde \Delta}|}{\sqrt{({\tilde p}_n)^2+|\Delta|^2}},
\end{eqnarray}
where $\Gamma_0=c_i\pi{N(0)}{\int}d\Omega_{\bf k'}|V_{\bf k_Fk_F'}|^2$. Self-consistency imposes $p_n/|\Delta|={\tilde p}_n/|{\tilde \Delta}|$, which ensures that the gap equation $|\Delta|=-gT\sum_{n,{\bf k}}{\rm Tr}[G(ip_n,{\bf k})\tau_1/2]$:
\begin{equation}
|\Delta|=gT\sum_{n}\int{d{\bf k}}\frac{|{\tilde \Delta}|}{\xi_{\bf k}^2+|{\tilde \Delta}|^2+({\tilde p}_n)^2}=\sum_n\frac{gTN(0)\pi}{\sqrt{1+({\tilde p}_n/|{\tilde \Delta}|)^2}}=\sum_n\frac{gTN(0)\pi}{\sqrt{1+({p}_n/|\Delta|)^2}},  
\end{equation}
reduces exactly to its clean form. This indicates that the gap equation is insensitive to non-magnetic impurities, i.e., the superconducting gap experiences a null renormalization from non-magnetic impurities~\cite{anderson1959theory,suhl1959impurity,skalski1964properties,andersen2020generalized}.  

In contrast, the superfluid density derives from a current–current correlation function, where vertex corrections due to impurity scattering necessarily appear, suppressing stiffness even when the gap is unaffected.

\end{appendix}
\end{widetext}

%
\end{document}